\documentclass[reqno,tbtags]{amsproc}
\usepackage{amssymb}
\usepackage{epsfig}
\usepackage{euscript}
\usepackage{eufrak}
\usepackage{fancybox}
\numberwithin{equation}{section} 
\newtheorem{theorem}{Theorem}[section]

\newtheorem{proposition}{Proposition}[section]
\theoremstyle{definition}
\newtheorem{definition}{Definition}[section]
\newtheorem{remark}{Remark}
\newtheorem{deficiency}{Deficiency}

\newcommand{\LunKonst}{\boldsymbol{C}}

\newcommand{\Ruat}{u_{\alpha,t}}
\newcommand{\RuaI}{u_{\alpha}}

\newcommand{\uatVaR}{u_{\alpha,t}^{\text{\tiny[{\sl VaR}]}}}

\def\paramT{\delta}

\renewcommand{\P}{\mathsf{P}}
\newcommand{\E}{\mathsf{E}\,}

\newcommand{\D}{\mathsf{D}\,}

\newcommand{\momgen}{M}

\newcommand{\Rline}{\mathsf{R}}

\newcommand{\UGauss}[2]{\varPhi_{\left({#1},{#2}\right)}}
\newcommand{\Ugauss}[2]{\varphi_{\left({#1},{#2}\right)}}

\newcommand{\BesselI}[1]{I_{#1}}
\newcommand{\Kummer}{U}

\newcommand{\probR}[1]{\boldsymbol{\psi}_{#1}}

\newcommand{\adjustL}{\varkappa}

\newcommand{\homN}[1]{N_{#1}}
\newcommand{\homV}[1]{V_{#1}}
\newcommand{\homR}[1]{R_{#1}}

\newcommand{\Ass}[1]{\bar{#1}}

\newcommand{\RtimeR}[1]{\Upsilon_{#1}}

\newcommand{\AInt}[2]{{\mathcal{#1}}_{#2}}

\newcommand{\Ss}[1]{S_{\!#1}}

\newcommand{\Y}[1]{Y_{#1}}
\newcommand{\T}[1]{T_{#1}}

\newcommand{\Span}{\Delta{c}}

\newcommand{\moplus}{m_{_{\scriptscriptstyle\vartriangle}}}
\newcommand{\mominus}{m_{_{\scriptscriptstyle\triangledown}}}
\newcommand{\Doplus}{D_{{\scriptscriptstyle\vartriangle}}}
\newcommand{\Dominus}{D_{{\scriptscriptstyle\triangledown}}}

\newcommand{\muIG}{\mu}
\newcommand{\HmuIG}{\hat{\mu}}
\newcommand{\lambdaIG}{\lambda}

\newcommand{\funU}[1]{\mathsf{z}_{#1}}

\newcommand{\NQuant}[1]{\kappa_{#1}}

\def\paramY{\rho}

\newcommand{\parA}[1]{a_{#1}}
\newcommand{\parB}[1]{b_{#1}}
\newcommand{\kumK}[1]{k_{#1}}
\newcommand{\kumL}[1]{l_{#1}}

\newcommand{\cS}{c^{\ast}}

\def\IME{\emph{In\-su\-ran\-ce: Ma\-the\-ma\-tics and Eco\-no\-mics\/}}
\def\DAN{\emph{Doklady Akademii Nauk\/}}
\def\GP{\emph{The Geneva Papers on Risk and Insurance -- Issues and Practice\/}}
\def\JIR{\emph{Journal of Insurance Regulation\/}}
\def\JACF{\emph{Journal of Applied Corporate Finance\/}}
\def\MF{\emph{Mathematical Finance\/}}
\def\RMIR{\emph{Risk Management and Insurance Review\/}}
\def\SAJ{\emph{Scan\-di\-na\-vian Ac\-tua\-rial Journal\/}}

\begin{document}

\author[Vsevolod K. Malinovskii]{Vsevolod K. Malinovskii\footnote{This work was
supported by RFBR (grant No.~19-01-00045).}}


\address{Central Economics and Mathematics Institute (CEMI) of Russian Academy of Science,
117418, Nakhimovskiy prosp., 47, Moscow, Russia}

\email{Vsevolod.Malinovskii@mail.ru, admin@actlab.ru}

\urladdr{http:/\!/www.actlab.ru}

\title[VALUE-AT-RISK SUBSTITUTE FOR NON-RUIN CAPITAL]{VALUE-AT-RISK SUBSTITUTE FOR
NON-RUIN \\[2pt] CAPITAL IS FALLACIOUS AND REDUNDANT}

\maketitle

\begin{abstract}
This seemed impossible to use a theoretically adequate but too sophisticated
risk measure called non-ruin capital, whence its widespread (including
regulatory documents) replacement with an inadequate, but simple risk measure
called Value-at-Risk. Conflicting with the idea by Albert Einstein that
``everything should be made as simple as possible, but not simpler'', this led
to fallacious, and even deceitful (but generally accepted) standards and
recommendations. Arguing from the standpoint of mathematical theory of risk, we
aim to break this impasse.
\end{abstract}

\vskip 18pt

\section{Introduction}\label{dgfjhngjk}

The basis of Solvency~II system (see Directives \cite{[Directive=2009]},
\cite{[Directive=2014]}) is the Value-at-Risk set as the main measure of risk.
Various criticisms (see, e.g., \cite{[Artzner=et=al.=1999]},
\cite{[Culp=1997]},  \cite{[Cummins=1994]}, \cite{[Doff=2008]},
\cite{[Eling=Schmeiser=2010]}, \cite{[Eling=Schmeiser=Schmit=2007]},
\cite{[Floreani=2013]}, \cite{[HeepAltiner=2018]}, \cite{[Marano=Siri=2017]},
\cite{[Mittnik=2011]}, \cite{[Pfeifer=Strassburger=2008]}) are directed against
this basis. Floreani (see \cite{[Floreani=2013]}) expressed it categorically:
\begin{quotation}
{\small the Solvency~II regime uses an inadequate risk measure to compute the
Solvency Capital Requirement $\dots$ The metric used by regulators, which is
based on a total risk measure such as the Value-at-Risk, is not a balanced
solution between effectiveness and simplicity, but is simply wrong and could
lead to significant adverse side effects, ultimately resulting in a generalized
European insurance industry crisis in the case of a hard market shortfall.}
\end{quotation}

Surprisingly, the text of Directives \cite{[Directive=2009]},
\cite{[Directive=2014]} contains clear evidence of controversy regarding risk
measures: the phrase\footnote{See Directive \cite{[Directive=2009]}, Article
101: Calculation of the Solvency Capital Requirement.} from Directive
\cite{[Directive=2009]} that the Solvency Capital Requirement (SCR) ``shall
correspond to the \emph{Value-at-Risk\/} of the basic own funds of an insurance
or reinsurance undertaking subject to a confidence level of 99{.}5\% over a
one-year period'' dramatically differs from the phrase in the same Directive
\cite{[Directive=2009]}, that\footnote{In the preambular paragraph (64) for
Directive \cite{[Directive=2009]}, it is said as follows: ``the Solvency
Capital Requirement should be determined as the economic capital to be held by
insurance and reinsurance undertakings \emph{in order to ensure that ruin
occurs} no more often than once in every 200 cases''.} the SCR determines the
economic capital which an insurance company must hold in order to guarantee a
\emph{one-year ruin probability\/} of at most 0{.}5\%.

Every risk theory expert knows that, dealing with solvency, it is appropriate
to address the occurrence of ruin within a year (and its probability), rather
than the capital deficit at the end of this year (and its probability).
Stepping back one step more, the aggregate claim amount distribution and the
probability of ruin within finite time, which are the basis for determining the
Value-at-risk and non-ruin capital respectively, were always clearly
distinguished in the risk theory.

In this paper, relying on inverse Gaussian approximations in the problem of
level crossing by a compound renewal processes and on associated results for
the level which a compound renewal process crosses with a given probability,
obtained in \cite{[Malinovskii=2018=dan=1]} and
\cite{[Malinovskii=2020=dan=2]}, we show that the non-ruin capital, being a
theoretically sound risk measure, is not inferior to the Value-at-risk in
simplicity even in a fairly general risk model.

The rest of this paper is arranged as follows. In Section~\ref{qaesryetj}, we
introduce the risk model, focussing on the difference between the Value-at-Risk
and non-ruin capital. In Section~\ref{sdrthyjhf}, using a series of well-known
results, we show that in the exponential risk model the analysis of non-ruin
capital is not much harder than the analysis of Value-at-Risk. In
Section~\ref{wa4e5yrtyr}, we show that in the general risk model most results
so much discussed in the literature are fit for the analysis of Value-at-Risk,
but not of non-ruin capital. We turn to recent advances in the direct and
inverse level crossing problems (see \cite{[Malinovskii=2018=dan=1]},
\cite{[Malinovskii=2020=dan=2]}) which are suitable for a deep insight into the
structure of non-ruin capital. In Section~\ref{ewrtyh6jh5rt}, we present
numerical results obtained by both analytical technique and direct simulation,
in order to illustrate the non-ruin capital's structure. The final conclusion
of this paper, given in Section~\ref{sartg54yh}, is that the analytical
structure of non-ruin capital is simple enough, and this measure of risk can be
used per se, without resorting to any substitute.

\section{Model and main definitions}\label{qaesryetj}

In the risk theory, the quantitative analysis is based on the annual\footnote{
This model, traditionally called Lundberg's collective risk model, is most
useful (see \cite{[Malinovskii=WS=1]}) as a building block for multi-year
models. From the angle of Directive \cite{[Directive=2009]}, this modeling is
very close to building an internal model.} model that formalizes the concept of
collective risk. Given that time is operational and $t$ is the length of the
year, for $0\leqslant s\leqslant t$ the claim arrival process is
\begin{equation}\label{srdtykuu}
\homN{s}=\max\Big\{n>0:\sum_{i=1}^{n}\T{i}\leqslant s\Big\},
\end{equation}
or $0$, if $\T{1}>s$, the cumulative claim payout process is
\begin{equation}\label{esrytehjethjr}
\homV{s}=\sum_{i=1}^{\homN{s}}\Y{i},
\end{equation}
or $0$, if $\T{1}>s$, and the balance of income and outcome is modeled by the
risk reserve process
\begin{equation}\label{567u67i76}
\homR{s}=u+cs-\homV{s},
\end{equation}
which starts at time zero at the point $u>0$, called initial capital. Here
$c\geqslant 0$ is called premium intensity (or, for the sake of brevity,
price), $\T{i}\overset{d}{=}T$, $i=1,2,\dots$, are i.i.d. intervals between
claims, and $\Y{i}\overset{d}{=}Y$, $i=1,2,\dots$, are i.i.d. claim sizes. It
is generally assumed that these sequences are independent of each other.

\begin{figure}[t]
\center{\includegraphics[scale=.8]{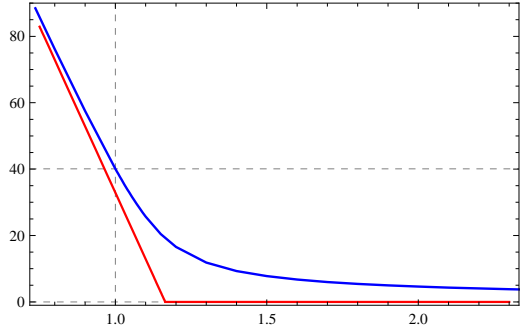}}
\caption{\small Graphs ($X$-axis is $c$) of $\uatVaR(c)$ (red) and $\Ruat(c)$
(blue), drawn for $T$ and $Y$ exponentially distributed with parameters
$\paramT=1$, $\paramY=1$, and $\alpha=0{.}05$, $t=200$. Horizontal grid line:
$\Ruat(\paramT/\paramY)=40{.}0844$. Vertical grid line:
$\paramT/\paramY=1$.}\label{dgjtkyasefrgew}
\end{figure}

In what follows, $\alpha$ is a reasonably small positive real number, e.g.,
$\alpha=0{.}05$.

\begin{definition}\label{aefgwegw}
The \emph{Value-at-Risk} $\uatVaR(c)$, $c\geqslant 0$, is a positive solution
to the equation
\begin{equation}\label{e5r6u7iytik}
\P\{\homR{t}<0\}=\alpha;
\end{equation}
for those $c$, for which this solution is negative, we set $\uatVaR(c)$ equal
to zero. The \emph{non-ruin capital} $\Ruat(c)$, $c\geqslant 0$, is a positive
solution to the equation
\begin{equation}\label{dstghjyrtjky}
\P\big\{\inf_{0\leqslant s\leqslant t}\homR{s}<0\big\}=\alpha;
\end{equation}
for those $c$, for which this solution is negative, we set $\Ruat(c)$ equal to
zero.
\end{definition}

Note that the left-hand side of \eqref{dstghjyrtjky} is the probability of ruin
within time $t$, i.e.,
\begin{equation*}
\begin{aligned}
\probR{t}(u,c)&=\P\big\{\inf_{0\leqslant s\leqslant t}\homR{s}<0\big\}
\\[0pt]
&=\P\{\RtimeR{u,c}\leqslant t\},
\end{aligned}
\end{equation*}
where $\RtimeR{u,c}=\inf\left\{s>0:\homV{s}-cs>u\right\}$, or $+\infty$, if
$\homV{s}-cs\leqslant u$ for all $s\geqslant 0$, is the time of the first ruin.
In these terms, equation \eqref{dstghjyrtjky} rewrites as
\begin{equation}\label{wertyjhft}
\P\{\RtimeR{u,c}\leqslant t\}=\alpha.
\end{equation}

Obviously, to investigate solvency in the usual sense of non-ruin\footnote{In
risk theory, the event of ruin is traditionally synonymous with bankruptcy, and
solvency is usually measured by the probability of ruin.}, we must focus on
$\Ruat(c)$, rather than on $\uatVaR(c)$. Since $\inf_{0\leqslant s\leqslant
t}\homR{s}$ is always less than or equal to $\homR{t}$, we have
\begin{equation}\label{sdtyhtfgj}
\uatVaR(c)\leqslant\Ruat(c),\quad c\geqslant 0,
\end{equation}
and $\uatVaR(c)$ always underestimates $\Ruat(c)$. The underestimating of
$\Ruat(c)$ by $\uatVaR(c)$ can be significant (see Fig.~\ref{dgjtkyasefrgew}).
To deal with this problem quantitatively, rather than qualitatively, we must
calculate both $\Ruat(c)$ and $\uatVaR(c)$ in the risk model
\eqref{srdtykuu}--\eqref{567u67i76}, striving for the most general assumptions
about $T$ and $Y$. Traditionally, it has been considered possible to achieve
success in this endeavor for $\uatVaR(c)$, but not for $\Ruat(c)$; our aim is
to break this impasse.

\begin{figure}[t]
\center{\includegraphics[scale=.8]{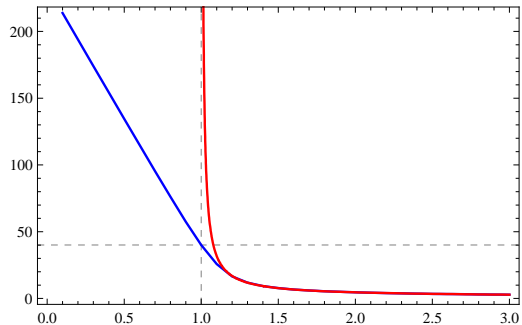}}
\caption{\small Graphs ($X$-axis is $c$) of $\Ruat(c)$ (blue) and $\RuaI(c)$
(red), drawn for $T$ and $Y$ exponentially distributed with parameters
$\paramT=1$, $\paramY=1$, and $\alpha=0{.}05$, $t=200$. Horizontal grid line:
$\Ruat(\cS)=40{.}08$. Vertical grid line: $\cS=1$.}\label{wqetyrjh}
\end{figure}

\begin{remark}\label{adsfgherjhr}
Let us write $\cS=\E{Y}/\E{T}$ and introduce the ultimate ruin probability
$\probR{\infty}(u,c)=\P\big\{\inf_{s\geqslant 0}\homR{s}<0\big\}$, which is
equivalently written as $\P\{\RtimeR{u,c}<\infty\}$. It has been studied in
detail (see, e.g., \cite{[Rolski=et=al.=1999]}). Plainly,
$\P\{\RtimeR{u,c}\leqslant t\}\leqslant\P\{\RtimeR{u,c}<\infty\}$.

In \cite{[Kalashnikov=1997]}, Chapter~6, Section~2, the insurer's risk is
measured by the ultimate ruin probability $\P\{\RtimeR{u,c}<\infty\}$ and the
``minimal admissible initial capital'' is introduced as a solution to the
equation
\begin{equation}\label{aesfgewge}
\P\{\RtimeR{u,c}<\infty\}=\alpha.
\end{equation}
In our notation, this is $\RuaI(c)$, $c>\cS$. Plainly (see
Fig.~\ref{wqetyrjh}), $\Ruat(c)\leqslant\RuaI(c)$, $c>\cS$, and $\RuaI(c)$ is
tending to infinity, as $c\to\cS$.

Assuming that the ``insurer wants to attract as many clients as possible
keeping the relative safety loading at the lowest possible level''
(\cite{[Kalashnikov=1997]}, pp.~172--173), in \cite{[Kalashnikov=1997]}
focussed is $\RuaI(c)$, as $c\to\cS$. Thus, the problem to explore ``the
initial capital securing a prescribed risk level when the relative safety
loading tends to zero'' (\cite{[Kalashnikov=1997]}, p.~27) is put forth.

In our opinion, the focus on $\RuaI(c)$, as $c\to\cS$, does not help the
insurer ``to attract as many clients as possible keeping the relative safety
loading at the lowest possible level'' (\cite{[Kalashnikov=1997]},
pp.~172--173), given that ``the insurer accepts at most $\alpha$ as an
acceptable risk level'' (\cite{[Kalashnikov=1997]}, p.~172). And even worse,
this is hard to accept that it ``can help the insurer to determine whether the
initial capital suffices to start the business'' (\cite{[Kalashnikov=1997]},
p.~175) because the theory developed in \cite{[Kalashnikov=1997]} claims that
when the insurer's price $c$ decreases to the equilibrium price $\cS$, what
often happens in some years of the real insurance business and what is far from
tragic, ``the initial capital securing a prescribed risk level'' is tending to
infinity.
\end{remark}

\section{Value-at-Risk and non-ruin capital in exponential case}\label{sdrthyjhf}

The additional assumption that $T$ and $Y$ in the model
\eqref{srdtykuu}--\eqref{567u67i76} are exponentially distributed with
parameters $\paramT$ and $\paramY$, yields many items of our interest in the
analytical form, in terms of elementary or special functions, such as modified
Bessel functions $\BesselI{k}(x)$, $x\geqslant 0$, of the first kind of order
$k$.

In what follows, we denote the cumulative distribution function (c.d.f.) of a
standard Gaussian distribution by $\UGauss{0}{1}(x)$, $x\in\Rline$. The
corresponding probability density function (p.d.f.) is denoted by
$\Ugauss{0}{1}(x)$, $x\in\Rline$. The $(1-\alpha)$-quantile of this
distribution is denoted by $\NQuant{\alpha}=\UGauss{0}{1}^{-1}(1-\alpha)$.
Plainly, $0<\NQuant{\alpha}<\NQuant{\alpha/2}$ for $0<\alpha<1/2$.

\subsection{Value-at-Risk and aggregate claim amount distribution}\label{asergerhye}

In the exponential case, for the aggregate claim amount $\homV{t}$ we have the
following widely known closed-form results:
\begin{equation}\label{sadfdhgdfjngf}
\E(\homV{t})=(\paramT/\paramY)\,t,\quad
\D(\homV{t})=2\,(\paramT/\paramY^{2})\,t,
\end{equation}
and
\begin{equation}\label{sdrttghkmg}
\begin{aligned}
\P\{\homV{t}\leqslant
x\}&=e^{-t\paramT}+e^{-t\paramT}\sum_{n=1}^{\infty}\frac{(t\paramT)^n}{n!}
\frac{\paramY^n}{\Gamma(n)}\int_{0}^{x}e^{-\paramY z}\,z^{n-1}\,dz
\\[0pt]
&=e^{-t\paramT}+e^{-t\paramT}\big(\paramT\paramY\,t\big)^{1/2}
\int_{0}^{x}z^{-1/2}\BesselI{1}\big(2\sqrt{\paramT\paramY\,
tz}\,\big)\,e^{-\paramY z}\,dz
\end{aligned}
\end{equation}
for $x>0$, and zero otherwise.

An important observation is that equation \eqref{e5r6u7iytik} rewrites as
\begin{equation}\label{wq4e5y65r}
\P\{\homV{t}>u+ct\}=\alpha,
\end{equation}
and its solution $\uatVaR(c)$ is (see, e.g., \cite{[Sandstrom=2011]},
Section~14.3.2) a percentile (or quantile) of the distribution of the aggregate
claim amount distribution at the year-end time point $t$.

Using equality \eqref{sdrttghkmg}, we express equation \eqref{wq4e5y65r} in a
closed form, whence $\uatVaR(c)$ is an implicit function defined by the
equation
\begin{equation}\label{srdthrjh}
e^{-t\paramT}+e^{-t\paramT}\big(\paramT\paramY\,
t\big)^{1/2}\int_{0}^{u+ct}z^{-1/2}\BesselI{1}\big(2\sqrt{\paramT\paramY\,
tz}\,\big)\,e^{-\paramY z}\,dz=1-\alpha.
\end{equation}
This implicit function can not be found in a closed form, but it can be
calculated numerically. The graph of $\uatVaR(c)$, $c\geqslant 0$, was drawn in
Fig.~\ref{dgjtkyasefrgew} in this way.

Since $\homV{t}$ is (see \eqref{esrytehjethjr}) the sum of $\homN{t}$ i.i.d.
random variables, where $\E(\homN{t})=\paramT t$, it seems natural to address
the asymptotic analysis of $\uatVaR(c)$, as $t\to\infty$. The assumption that
$t$ is large, which allows us to turn to the central limit theory, is sensible
in terms of applications for the following reasons: time in the model
\eqref{srdtykuu}--\eqref{567u67i76} is operational (see, e.g.,
\cite{[Sparre=Andersen=1957]} p.~219), rather than calendar. This time,
measured in monetary units, is proportional to the ball-park figure of the
annual financial transactions of the company. Consequently, the assumption that
$t\to\infty$ means that this ball-park figure is large, i.e., the insurer's
portfolio size is large.

Since $\homV{t}$ is asymptotically normal with mean and variance given in
\eqref{sadfdhgdfjngf}, equation \eqref{srdthrjh} is closely related to the
equation
\begin{equation}\label{re6u756iu5}
\UGauss{0}{1}\left(\frac{u+ct-(\paramT/\paramY)\,t}
{\sqrt{2\,(\paramT/\paramY^{\,2})\,t}}\right)=1-\alpha,
\end{equation}
whose solution
$(\paramT/\paramY-c)\,t+\big(\sqrt{2\paramT}/\paramY\big)\,\NQuant{\alpha}\sqrt{t}$
is straightforward. Applying simple arguments based on the proximity of two
implicit functions, we conclude that for all $c\geqslant 0$
\begin{equation}\label{wer5t75iry}
\uatVaR(c)=\max\Big\{0,\big(\paramT/\paramY-c\big)\,t
+\frac{\sqrt{2\paramT}}{\paramY}\,\NQuant{\alpha}\sqrt{t}\,(1+{o}(1))\Big\},\quad
t\to\infty.
\end{equation}

\subsection{Probability of ruin and non-ruin capital}\label{qewrtehr}

In the exponential case, we have the following widely known (see, e.g.,
\cite{[Malinovskii=1998]}, Remark~2) closed-form result:
\begin{equation}\label{yhukgykhkh}
\P\{\RtimeR{u,c}\leqslant
t\}=\P\{\RtimeR{u,c}<\infty\}-\frac{1}{\pi}\int_0^{\pi}f(x)\,dx,
\end{equation}
where
\begin{equation*}
\P\{\RtimeR{u,c}<\infty\}=\begin{cases}1,&\paramT/(c\paramY)\geqslant 1,
\\[4pt]
\dfrac{\paramT}{c\paramY}\,\exp\{-u\,(c\paramY-\paramT)/c\},&\paramT/(c\paramY)<1,
\end{cases}
\end{equation*}
and
\begin{equation*}
\begin{aligned}
f(x)&=(\paramT/(c\paramY))\big(1+\paramT/(c\paramY)-2\sqrt{\paramT/(c\paramY)}\cos
x\big)^{-1}
\\[0pt]
&\times\exp\Big\{u\paramY\,\big(\sqrt{\paramT/(c\paramY)}\cos x-1\big)-t\paramT
(c\paramY/\paramT)
\\[0pt]
&\times\big(1+\paramT/(c\paramY)-2\sqrt{\paramT/(c\paramY)}\cos x\big)\Big\}
\\[0pt]
&\times\big(\cos\big(u\paramY\sqrt{\paramT/(c\paramY)}\sin x\big)
-\cos\big(u\paramY\sqrt{\paramT/(c\paramY)}\sin x+2x\big)\big).
\end{aligned}
\end{equation*}

Using equality \eqref{yhukgykhkh}, we express the left-hand side of equation
\eqref{wertyjhft} in a closed form, whence $\Ruat(c)$ is an implicit function
defined by this equation. The same as $\uatVaR(c)$, this implicit function can
not be found in a closed form, but can be calculated numerically. The graph of
$\Ruat(c)$, $c\geqslant 0$, was drawn in Fig.~\ref{dgjtkyasefrgew} in this way.

\begin{figure}[t]
\center{\includegraphics[scale=.8]{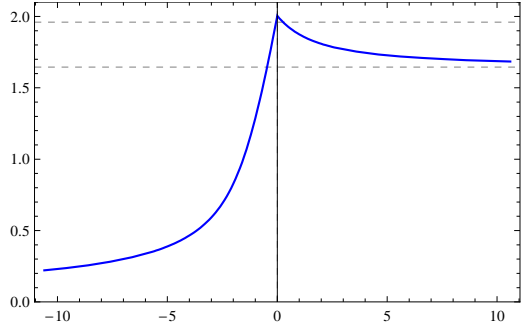}}
\caption{\small Graph ($X$-axis is $x$) of $\funU{\alpha,t}(x)$, drawn for $T$
and $Y$ exponentially distributed with parameters $\paramY=1$, $\paramT=1$, and
$\alpha=0{.}05$, $t=200$. Horizontal grid lines: $\NQuant{\alpha}=1{.}645$ and
$\NQuant{\alpha/2}=1{.}960$.}\label{serfgsegsrdXX}
\end{figure}

The function $\Ruat(c)$, $c\geqslant 0$, can be analyzed asymptotically (see,
e.g., \cite{[Malinovskii=2012]}, Theorems~3.1 and~3.2; this analysis was based
on the properties of Bessel functions). In particular, we have (see
\cite{[Malinovskii=2012]}, Theorem 3.2)
\begin{equation}\label{dtyrkkjuy}
\Ruat(c)=\begin{cases}\left(\paramT/\paramY-c\right)t
+\dfrac{\sqrt{2\,\paramT}}{\paramY}\,\funU{\alpha,t}
\bigg(\dfrac{\paramY\,\big(\paramT/\paramY-c\big)}{\sqrt{2\,\paramT}}\sqrt{t}\,\bigg)\sqrt{t},&0\leqslant
c\leqslant\cS,
\\[4pt]
\dfrac{\sqrt{2\,\paramT}}{\paramY}\,\funU{\alpha,t}
\bigg(\dfrac{\paramY\,\big(\paramT/\paramY-c\big)}{\sqrt{2\,\paramT}}\sqrt{t}\,\bigg)\sqrt{t},&c>\cS,
\end{cases}
\end{equation}
where $\cS=\paramT/\paramY$ and the function $\funU{\alpha,t}(x)$,
$x\in\Rline$, is (see Fig.~\ref{serfgsegsrdXX}) continuous, monotone
increasing, as $x$ increases from $-\infty$ to $0$, monotone decreasing, as $x$
increases from $0$ to $\infty$, and such that
$\lim_{x\to-\infty}\funU{\alpha,t}(x)=0$,
$\lim_{x\to\infty}\funU{\alpha,t}(x)=\NQuant{\alpha}$, and
$\funU{\alpha,t}(0)=\NQuant{\alpha/2}(1+{o}(1))$, as $t\to\infty$. In
particular, we have
\begin{equation}\label{sfddfdbcsbb}
\begin{aligned}
\Ruat(0)&=({\paramT}/{\paramY})\,t+\frac{\sqrt{2\paramT}}{\paramY}
\,\NQuant{\alpha}\sqrt{t}\,(1+{o}(1)),\quad t\to\infty,
\\[-4pt]
\Ruat(\cS)&=\frac{\sqrt{2\paramT}}{\paramY}\,\NQuant{\alpha/2}\sqrt{t}\,(1+{o}(1)),\quad
t\to\infty.
\end{aligned}
\end{equation}
First equality in \eqref{sfddfdbcsbb} is straightforward; see, e.g., (2.4) in
\cite{[Malinovskii=Kosova=2014]}. Second equality in \eqref{sfddfdbcsbb} is
Theorem 3.1 in \cite{[Malinovskii=2012]}.

\section{Value-at-Risk and non-ruin capital in general case}\label{wa4e5yrtyr}

It is widely believed that in the general case, the situation described in
Section~\ref{sdrthyjhf} deteriorates dramatically, and the non-ruin capital
becomes intractable. First, we clarify the reasons for this belief. Second, we
show that the situation in the general case is not so bad due to several
innovative approaches.

\subsection{Value-at-Risk and aggregate claim amount distribution}\label{tfuyrtiyfi}

In the general case, there is no hope to get explicit equalities like
\eqref{sadfdhgdfjngf} or \eqref{sdrttghkmg} for all $t$. But since the
asymptotic analysis, as $t\to\infty$, is based on the fairly general central
limit theory, it is easy to obtain analogues for \eqref{re6u756iu5} and
\eqref{wer5t75iry}. To be specific, $\P\{\homV{t}\leqslant x\}$ is approximated
by $\UGauss{M_{V}t}{D_{V}t}(x)$, as $t\to\infty$, where\footnote{It is
noteworthy that
$\E(\homN{t})=M_{N}\,t+\frac{\D{T}-(\E{T})^{2}}{2\,(\E{T})^{2}}+{o}(1)$,
$\D(\homN{t})=D_{N}^{\,2}\,t+{o}(t)$,
$\E(\homV{t})=M_{V}\,t+\frac{\E{Y}(\D{T}-(\E{T})^{2})}{2\,(\E{T})^{2}}+{o}(1)$,
$\D(\homV{t})=D_{V}^{\,2}\,t+{o}(t)$, $t\to\infty$.}
\begin{equation*}
\begin{aligned}
M_{N}&= 1/\E{T},&M_{V}&=\E{Y}/\E{T},
\\[0pt]
D_{N}^{\,2}&=\D{T}/(\E{T})^{3},&
D_{V}^{\,2}&=\E(T\E{Y}-Y\E{T})^{2}/(\E{T})^{3}.
\end{aligned}
\end{equation*}
This approximation, being a version of the central limit theorem, is valid
under well-known mild technical condition on $T$ and $Y$ and can be applied to
\eqref{wq4e5y65r}.

Therefore, though in the general case equation \eqref{wq4e5y65r} cannot be
written in terms of elementary or special functions, as it was done (see
\eqref{srdthrjh}) in the exponential case, for $t$ sufficiently large
\eqref{wq4e5y65r} is close to the equation (cf. \eqref{re6u756iu5})
\begin{equation}\label{zdxfvsdb}
\UGauss{0}{1}\left(\frac{u+ct-M_{V}t}{D_{V}\sqrt{t}}\right)=1-\alpha,
\end{equation}
whose closed-form solution $(M_{V}-c)\,t+\NQuant{\alpha}\,D_{V}\sqrt{t}$ is
straightforward. Applying simple arguments based on the proximity of two
implicit functions, we conclude that (cf. \eqref{wer5t75iry}) for all
$c\geqslant 0$
\begin{equation}\label{dfyujykguk}
\uatVaR(c)=\max\Big\{0,(M_{V}-c)\,t
+\NQuant{\alpha}\,D_{V}\sqrt{t}\,(1+{o}(1))\Big\},\quad t\to\infty.
\end{equation}
It is easy to see that the analysis in the general case differs a little,
regarding both applied technique and results, from the analysis in the
exponential case.

\subsection{Standard results for ruin probability}\label{fdghfgjgjh}

In the general case (except for some very special subcases), there is no hope
to express $\P\{\RtimeR{u,c}\leqslant t\}$ in terms of elementary or special
functions for all $t$. There is even less hope of finding in a closed form for
all $t$ the implicit function $\Ruat(c)$, $c\geqslant 0$, defined by the
corresponding equation \eqref{wertyjhft}, even if its left-hand side could be
represented in such terms.

Moreover, the results of asymptotic analysis, as $u\to\infty$, so much
discussed in the literature, are unsatisfactory from the angle of their further
application to asymptotical, as $t\to\infty$, analysis of the non-ruin capital
$\Ruat(c)$, $c\geqslant 0$. We will show this by referring to the normal (or
Cram{\'e}r's) and diffusion approximations that are best known. We start with
the former\footnote{What is said below about this approximation is folklore of
the risk theory and can be found in many standard textbooks, e.g., in
\cite{[Rolski=et=al.=1999]}.} and point out its deficiencies.

\subsubsection{Normal approximation}\label{dstyjtrjkty}

The primary assumption is that there exists a positive solution $\adjustL$,
called adjustment coefficient, to the equation (w.r.t. $r$)
\begin{equation}\label{sdfgrhjrgde}
\momgen_{X}(r)=1,
\end{equation}
called Lundberg's equation. Here $\momgen_{X}(r)=\E(e^{rX})$ is the moment
generating function of $X\overset{d}{=}Y-c\,T$; plainly, $\momgen_{X}(0)=1$.
This assumption is a significant limitation of the model. It implies that
$\momgen_{X}(r)$ has to exist in a neighborhood of $0$ or, in other words, that
the right tail of c.d.f. $F_{X}$ is exponentially bounded above. The latter
follows from Markov's inequality
\begin{equation}\label{xvbdnbsd}
1-F_{X}\left(x\right)\leqslant e^{-\adjustL x}\,\E(e^{\adjustL X})=e^{-\adjustL
x},\quad x>0.
\end{equation}

Starting with c.d.f. $F_{XT}(x,t)=\P\{X\leqslant x,T\leqslant t\}$ and having
$\adjustL>0$ found, we introduce the associated joint
distribution\footnote{See, e.g., Example~(b) in \cite{[Feller=1971]}, Chapter
XII, Section~4.}, whose c.d.f. $F_{\Ass{X}\Ass{T}}(x,t)=\P\{\Ass{X}\leqslant
x,\Ass{T}\leqslant t\}$ is defined by the equality\footnote{Commonly used
shorthand notation for it is $F_{\Ass{X}\Ass{T}}(dx,dt)=e^{\adjustL
z}F_{XT}(dx,dt)$.}
\begin{equation*}
F_{\Ass{X}\Ass{T}}\left(x,t\right)=\int_{-ct}^{x}\int_{0}^{t}e^{\adjustL
z}\,F_{XT}(dz,dw).
\end{equation*}
Plainly, this is a proper probability distribution.

Recall that $\cS=\E{Y}/\E{T}$. The normal (or Cram{\'e}r's) approximation is
formulated separately for $0\leqslant c<\cS$ and for $c>\cS$, with the case
$c=\cS$ excluded. For\, $0\leqslant c<\cS$, i.e., for\,
$\E{X}=\E{Y}-c\,\E{T}>0$, we write
\begin{equation*}
\mominus=\E{T}/\E{X},\quad\Dominus^{\,2}=\E(X\E{T}-T\E{X})^{2}/(\E{X})^{3}.
\end{equation*}
Plainly, we have $\mominus>0$ and $\Dominus^{\,2}>0$.

\begin{proposition}[Case $0\leqslant c<\cS$]\label{sdtghjrtyjkt}
Assume that\, p.d.f. of the random vector $(T,Y)$ is bounded above by a finite
constant and $0<\Dominus^{\,2}<\infty$. Then
\begin{equation*}
d_{u}=\sup_{\,t>0}\,\big|\,\P\{\RtimeR{u,c}\leqslant t\}-\UGauss{\mominus
u}{\Dominus^{\,2}u}(t)\,\big|\,={o}(1),\quad u\to\infty.
\end{equation*}
If, in addition, $\E(Y^{3})<\infty$, $\E(T^{3})<\infty$, then
$d_{u}={O}(u^{-1/2})$, as $u\to\infty$.
\end{proposition}

For $c>\cS$, i.e., for\, $\E{X}=\E{Y}-c\,\E{T}<0$, we write
\begin{equation*}
\begin{gathered}
\moplus=\E\Ass{T}/\E\Ass{X},\quad
\Doplus^{\,2}=\E(\Ass{X}\E\Ass{T}-\Ass{T}\E\Ass{X})^{2}/(\E\Ass{X})^{3},
\\[0pt]
\LunKonst=\frac{1}{\adjustL\,\E\Ass{X}}\exp\Big\{-\sum^{\infty}_{n=1}\frac{1}{n}\,
\P\{\Ss{n}>0\}-\sum^{\infty}_{n=1}\frac{1}{n}\,\P\{\Ass{S}_{n}\leqslant
0\}\Big\},
\end{gathered}
\end{equation*}
where $\Ass{X}_{i}\overset{d}{=}\Ass{X}$, $i=1,2,\dots$, and
$\Ass{T}_{i}\overset{d}{=}\Ass{T}$, $i=1,2,\dots$, are associated random
variables, and $\Ass{S}_{n}=\sum_{i=1}^n\Ass{X}_{i}$, and
$\Ass{Z}_{n}=\sum_{i=1}^{n}\Ass{T}_{i}$, $n=1,2,\dots$, are associated random
walks.

\begin{proposition}[Case $c>\cS$]\label{sdgehtdrth}
Assume that a solution $\adjustL>0$ to equation \eqref{sdfgrhjrgde} exists,
p.d.f. of the random vector $(T,Y)$ is bounded above by a finite constant, and
$0<\Doplus^{\,2}<\infty$. Then
\begin{equation*}
d_{u}=\sup_{\,t>0}\,\big|\,e^{\adjustL u}\,\P\{\RtimeR{u,c}\leqslant
t\}-\LunKonst\,\UGauss{\moplus u}{\Doplus^{\,2} u} (t)\,\big|={o}(1),\quad
u\to\infty.
\end{equation*}
If, in addition, $\E(T^{3})<\infty$, then $d_{u}={O}(u^{-1/2})$, as
$u\to\infty$.
\end{proposition}

In the exponential case, when $T$ and $Y$ are exponentially distributed with
parameters $\paramT>0$ and $\paramY>0$, straightforward calculations (see
\cite{[Malinovskii=Kosova=2014]}, Proposition 2.3) yield $\cS=\paramT/\paramY$,
\begin{equation}\label{srdthgtjrf}
\begin{aligned}
\LunKonst&=\paramT/(c\paramY),&\adjustL&=\paramY\,(1-\paramT/(c\paramY)),
\\[0pt]
\mominus&=-\frac{1}{c\,(1-\paramT/(c\paramY))},&
\Dominus^{\,2}&=-\frac{2\,(\paramT/(c\paramY))}{c^{2}\paramY\,(1-\paramT/(c\paramY))^{\,3}},
\\[0pt]
\moplus&=\frac{\paramT/(c\paramY)}{c\,(1-\paramT/(c\paramY))},&
\Doplus^{\,2}&=\frac{2\,(\paramT/(c\paramY))}{c^{2}\paramY\,(1-\paramT/(c\paramY))^{\,3}},
\end{aligned}
\end{equation}
and Propositions~\ref{sdtghjrtyjkt} and \ref{sdgehtdrth} are fused together, as
follows.

\begin{figure}[t]
\center{\includegraphics[scale=.8]{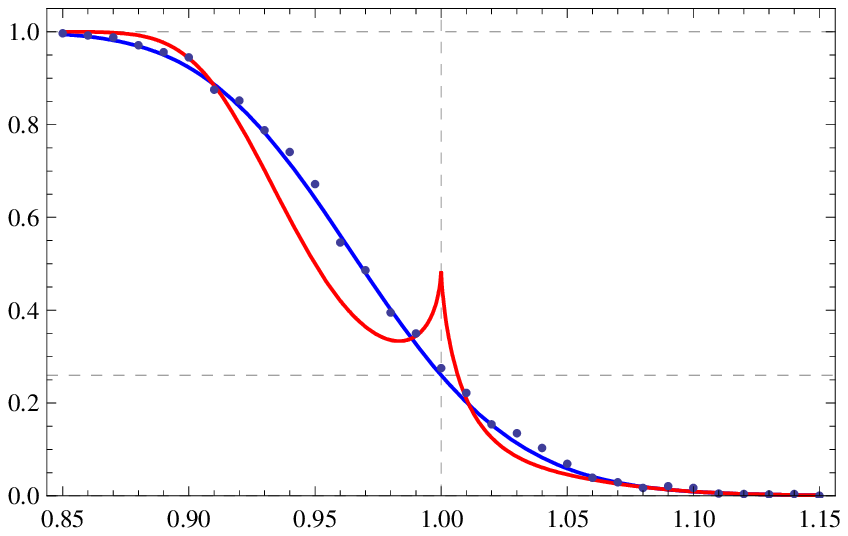}}
\caption{\small Graphs ($X$-axis is $c$) of $\P\{\RtimeR{u,c}\leqslant t\}$
(blue), of the approximations of Proposition~\ref{etrjykih} (red), and of
simulated values ($\Span=0{.}05$, $N=1000$) of $\P\{\RtimeR{u,c}\leqslant t\}$,
drawn for $T$ and $Y$ exponentially distributed with parameters $\paramY=1$,
$\paramT=1$, and $t=1000$, $u=50$. Horizontal grid line:
$\P\{\RtimeR{u,\cS}\leqslant t\}=0{.}26$.}\label{ewrtyhrtjr}
\end{figure}

\begin{figure}[t]
\center{\includegraphics[scale=.8]{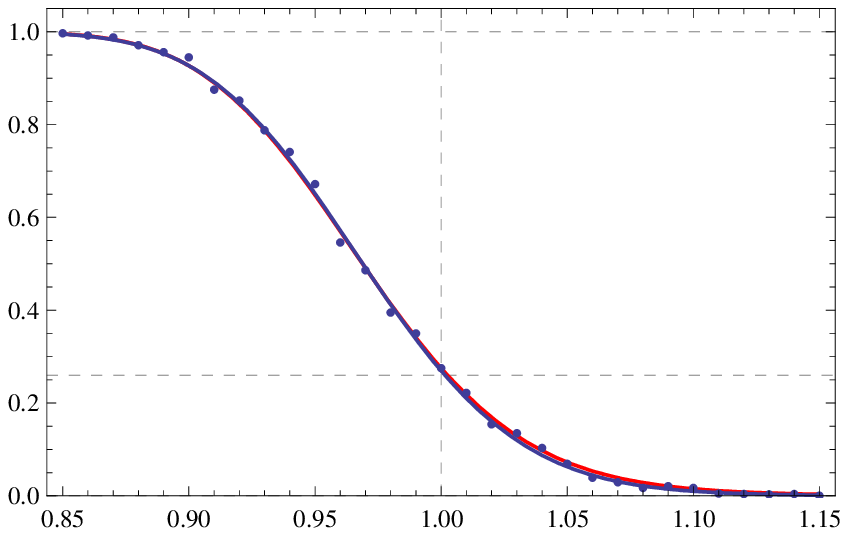}}
\caption{\small Graphs ($X$-axis is $c$) of $\P\{\RtimeR{u,c}\leqslant t\}$
(blue), of $\AInt{M}{u,c}(t)$ (red), and of simulated values ($\Span=0{.}05$,
$N=1000$) of $\P\{\RtimeR{u,c}\leqslant t\}$, drawn for $T$ and $Y$
exponentially distributed with parameters $\paramY=1$, $\paramT=1$, and
$t=1000$, $u=50$. Horizontal grid line: $\P\{\RtimeR{u,\cS}\leqslant
t\}=0{.}26$.}\label{sdrtfrjmty}
\end{figure}

\begin{proposition}\label{etrjykih}
In the renewal model with $T$ and $Y$ exponentially distributed with parameters
$\paramT>0$ and $\paramY>0$, we have for $0\leqslant c<\cS$
\begin{equation*}
\sup_{\,t>0}\,\left|\,\P\{\RtimeR{u,c}\leqslant t\}-\UGauss{\mominus
u}{\Dominus^{\,2} u}(t)\,\right|\,={o}(1),\quad u\to\infty,
\end{equation*}
where $\mominus>0$, $\Dominus^{\,2}>0$ are defined in \eqref{srdthgtjrf}, and
for $c>\cS$
\begin{equation*}
\sup_{\,t>0}\,\left|\,e^{\adjustL u}\,\P\{\RtimeR{u,c}\leqslant
t\}-\LunKonst\,\UGauss{\moplus u}{\Doplus^{\,2} u} (t)\,\right|\,={o}(1),\quad
u\to\infty,
\end{equation*}
where $\adjustL>0$, and $0<\LunKonst<1$, $\moplus>0$, $\Doplus^{\,2}>0$ are
defined in \eqref{srdthgtjrf}.
\end{proposition}

These results, if used for an asymptotic analysis of non-ruin capital,
$\Ruat(c)$, $c\geqslant 0$, have several deficiencies.

\begin{deficiency}[Limited applicability]\label{sarfhgrtj}
Proposition~\ref{sdgehtdrth} requires restrictive technical condition: $Y$ must
be (see \eqref{xvbdnbsd}) light-tailed.
\end{deficiency}

\begin{deficiency}[Flaw near $c=\cS$]\label{srdtyjhtykj}
Besides the fact that the case $c=\cS$ is formally excluded, the normal (or
Cram{\'e}r's) approximation fails for $c$ in a neighborhood of $\cS$. This is
illustrated in Fig.~\ref{ewrtyhrtjr}.
\end{deficiency}

\begin{deficiency}[Structural imbalance]\label{qwergghsdfh}
The structure of the approximation in Propositions~\ref{sdtghjrtyjkt} and
\ref{sdgehtdrth}  is significantly different from the structure of CLT--type
approximation used to get \eqref{zdxfvsdb} and \eqref{dfyujykguk}. This is
particularly evident when $c=0$, i.e., when the left-hand sides of equations
\eqref{e5r6u7iytik} and \eqref{wertyjhft} are the same and can be written as
$\P\{\homV{t}>u\}$. It is clear that, being solutions to the same equation,
$\Ruat(0)$ and $\uatVaR(0)$ coincide with each other. The CLT--type
approximation is
\begin{equation}\label{dtyujtyk}
\P\{\homV{t}>u\}\approx
\UGauss{0}{1}\left(\frac{u-M_{V}\big|_{\,c=0}\,t}{D_{V}\big|_{\,c=0}\sqrt{t}}\right),\quad
t\to\infty,
\end{equation}
whereas the approximation in Proposition~\ref{sdtghjrtyjkt} is
\begin{equation}\label{wetuyur}
\P\{\homV{t}>u\}\approx\UGauss{0}{1}\left(\frac{t-\mominus\big|_{\,c=0}\,
u}{\Dominus\big|_{\,c=0}\sqrt{u}}\right),\quad u\to\infty.
\end{equation}
When $T$ and $Y$ are exponentially distributed with parameters $\paramT$ and
$\paramY$, $\mominus\big|_{\,c=0}={\paramY}/{\paramT}$,
$\Dominus^{\,2}\big|_{\,c=0}={2\paramY}/{\paramT^2}$ in \eqref{dtyujtyk}, and
$M_{V}\big|_{\,c=0}={\paramT}/{\paramY}$,
$D_{V}^{\,2}\big|_{\,c=0}={2\paramT}/{\paramY^2}$ in \eqref{wetuyur}.

It is worth noting that if $\Ruat(0)$ would not tend to infinity, as
$t\to\infty$, the approximation \eqref{wetuyur} (unlike \eqref{dtyujtyk}) would
be useless to get information about $\Ruat(0)$, as $t\to\infty$. Fortunately,
$\Ruat(0)\sim M_{V}\,|_{\,c=0}\,t$, which tends to infinity, as $t\to\infty$.
Indeed, $\Ruat(0)$ is equal to $\uatVaR(0)$ and (see \eqref{dfyujykguk})
\begin{equation*}
\uatVaR(0)=M_{V}\big|_{\,c=0}\,t
+\NQuant{\alpha}\,D_{V}\big|_{\,c=0}\sqrt{t}\,(1+{o}(1)),\quad t\to\infty.
\end{equation*}

For $c$ sufficiently larger than $\cS$, i.e., for $c>K\cS$ with $K>1$
sufficiently large, $\Ruat(c)$ is finite regardless of $t$. Therefore, the
structural difference between the approximations of
Propositions~\ref{sdtghjrtyjkt} and \ref{sdgehtdrth} matters.
\end{deficiency}

\subsubsection{Diffusion approximation}\label{wsrdthytrj}

Because of the space limitations, we describe the deficiencies of simple and
corrected diffusion approximations for $\P\{\RtimeR{u,c}\leqslant t\}$
verbally. The idea of a simple diffusion approximation\footnote{What is said
below about this approximation is folklore of the risk theory and can be found
in many standard textbooks, e.g., in \cite{[Rolski=et=al.=1999]}.} is that the
original risk reserve process \eqref{567u67i76} has some similarities
(regarding the properties of distributions, rather than trajectories) with the
diffusion process, although its trajectories are continuous. Matching the
original and the auxiliary diffusion processes, one finds\footnote{Using (see,
e.g.,~\cite{[Billingsley=1999]}) Donsker's theorem, called also
Donsker--Prohorov's invariance principle.} that the distribution of the first
level crossing time for the former process is approximated by the distribution
of the first level crossing time for the latter process. This observation is
productive because the first passage probabilities in the diffusion model are
found in a closed form.

The diffusion process is skip-free; the idea behind the simple diffusion
approximation ignores the presence of overshoot in the original process with
discontinuous trajectories. The corrected diffusion approximation takes into
account this and other similar features of the initial process.

Congenital deficiency of both simple and adjusted diffusion approximations is
Deficiency~\ref{sarfhgrtj}: these results often require that the distribution
of $Y$ has a light tail. In addition, these results are valid under the
assumption\footnote{Often formulated as ``the safety loading is small and
positive''; it is often added that this is just the same as the heavy traffic
in the queuing theory.} that $c\to\cS+0$. Such regime is certainly a structural
drawback; this impedes the analysis of $\Ruat(c)$, $c\geqslant 0$.

\subsection{Innovative results for ruin probability}\label{fdghfgjgjh}

\begin{figure}[t]
\center{\includegraphics[scale=.8]{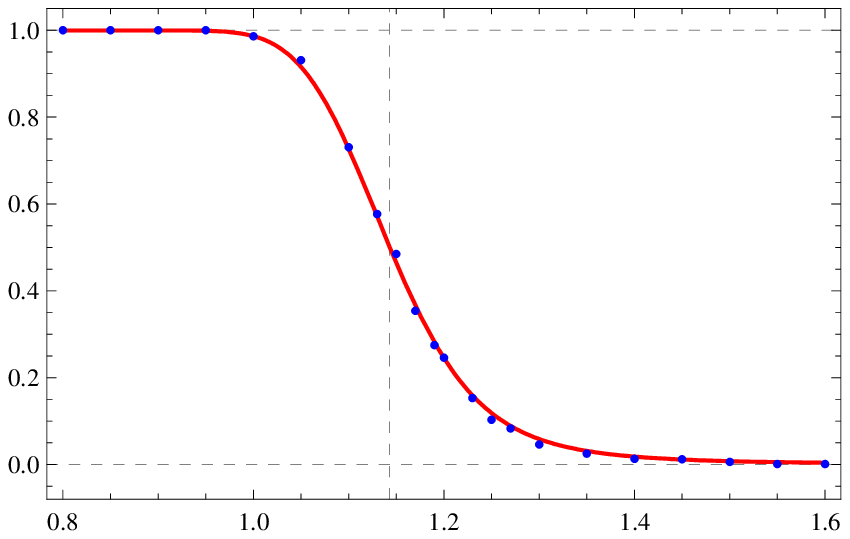}}
\caption{\small Graph ($X$-axis is $c$) of $\AInt{M}{u,c}(t)$ (red) and of
simulated values ($\Span=0{.}05$, $N=1000$) of $\P\{\RtimeR{u,c}\leqslant t\}$
(blue), drawn for $T$ which is 2-mixture with parameters $\paramT_1=1$,
$\paramT_2=2$, $p=2/3$, $Y$ which is Pareto with parameters $\parA{Y}=4{.}0$,
$\parB{Y}=0{.}35$, and $t=1000$, $u=40$.}\label{qwertyrjty}
\end{figure}

\begin{figure}[t]
\center{\includegraphics[scale=.8]{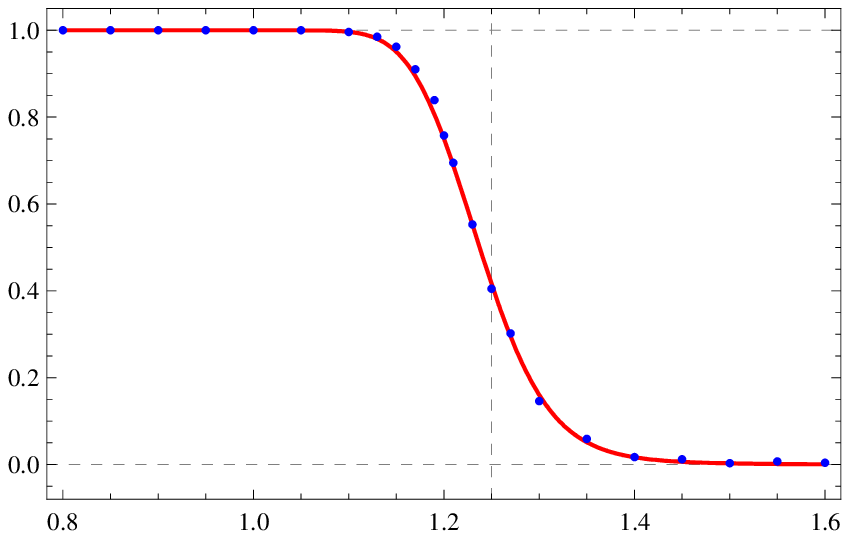}}
\caption{\small Graph ($X$-axis is $c$) of $\AInt{M}{u,c}(t)$ (red) and of
simulated values ($\Span=0{.}05$, $N=1000$) of $\P\{\RtimeR{u,c}\leqslant t\}$
(blue), drawn for $T$ which is Erlang with parameters $\paramT=6{.}0$, $k=4$,
$Y$ which is Pareto with parameters $\parA{Y}=4{.}0$, $\parB{Y}=0{.}4$, and
$t=1000$, $u=40$.}\label{ftuytrjityy}
\end{figure}

\begin{figure}[t]
\center{\includegraphics[scale=.8]{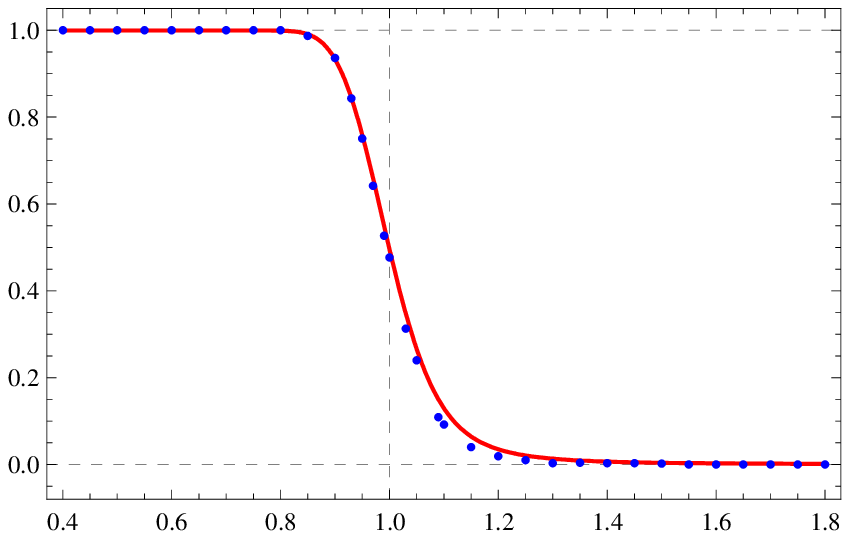}}
\caption{\small Graph ($X$-axis is $c$) of $\AInt{M}{u,c}(t)$ (red) and of
simulated values ($\Span=0{.}05$, $N=1000$) of $\P\{\RtimeR{u,c}\leqslant t\}$
(blue), drawn for $T$ which is Pareto with parameters $\parA{T}=4{.}0$,
$\parB{T}=0{.}4$. $Y$ which is Pareto with parameters $\parA{Y}=4{.}0$,
$\parB{Y}=0{.}4$, and $t=1000$, $u=40$.}\label{srtyikyer}
\end{figure}

For
\begin{equation}\label{adsfgherfjnfnm}
M=\E{T}/\E{Y},\quad D^{\,2}=((\E{T})^{2}\D{Y}+(\E{Y})^{2}\D{T})/(\E{Y})^{3},
\end{equation}
we write
\begin{equation}\label{srdfdsfgs}
\AInt{M}{u,c}(t)=\int_{0}^{\frac{ct}{u}}\frac{1}{x+1}\,
\Ugauss{cM(x+1)}{\frac{c^{2}D^{\,2}}{u}(x+1)}(x)\,dx.
\end{equation}

Bearing in mind that $\cS=\E{Y}/\E{T}$ equals to $1/M$ and denoting p.d.f. of
inverse Gaussian distribution by
\begin{equation*}
F(x;\muIG,\lambdaIG)=
\UGauss{0}{1}\bigg(\sqrt{\frac{\lambdaIG}{x}}\,\bigg(\frac{x}{\muIG}-1\bigg)\bigg)
+\exp\bigg\{\frac{2\lambdaIG}{\muIG}\bigg\}
\,\UGauss{0}{1}\bigg(-\sqrt{\frac{\lambdaIG}{x}}\,\bigg(\frac{x}{\muIG}+1\bigg)\bigg),\quad
x>0,
\end{equation*}
we can show by elementary calculations that
\begin{equation*}
\AInt{M}{u,c}(t)=\begin{cases}\Big(F\Big(\dfrac{ct}{u}+1;\muIG,\lambdaIG\Big)
-F\big(1;\muIG,\lambdaIG\big)
\Big)\,\Big|_{\,\muIG=\frac{1}{1-cM},\lambdaIG=\frac{u}{c^{\,2}D^{\,2}}}, &
0<c\leqslant\cS,
\\[12pt]
\exp\Big\{-\dfrac{2\lambdaIG}{\HmuIG}\Big\}\,\Big(F\Big(\dfrac{ct}{u}+1;
\HmuIG,\lambdaIG\Big)-F\big(1;\HmuIG,\lambdaIG\big)
\Big)\,\Big|_{\,\HmuIG=\frac{1}{cM-1},\lambdaIG=\frac{u}{c^{\,2}D^{\,2}}}, &
c>\cS.
\end{cases}
\end{equation*}

The following theorem, as well as its refinements like Edgeworth expansions
(see \cite{[Malinovskii=2018=dan=1]} and \cite{[M1]}, \cite{[M2]},
\cite{[M3]}), is called the inverse Gaussian approximation for
$\P\{\RtimeR{u,c}\leqslant t\}$.

\begin{theorem}\label{esrytrf}
Assume that p.d.f. $f_{T}(x)$ and $f_{Y}(x)$ are bounded above by a finite
constant, $D^{\,2}>0$, $\E({T}^{3})<\infty$, $\E({Y}^{3})<\infty$. Then for any
$c\geqslant 0$
\begin{equation*}
\sup_{t>0}\,\left|\,\P\{\RtimeR{u,c}\leqslant t\}
-\AInt{M}{u,c}(t)\,\right|={o}(1),\quad t,u\to\infty.
\end{equation*}
\end{theorem}

In a nutshell, the proof of Theorem~\ref{esrytrf} is based on Kendall's
identity which represents the first level crossing time's distribution in terms
of the convolution powers of p.d.f. $f_{T}(x)$ and $f_{Y}(x)$; then the
well-developed central limit theory is applied to these convolution powers.

\begin{remark}\label{asertehe}
The inverse Gaussian distribution\footnote{Theorem~\ref{esrytrf} shows that it
is central in the following approximation, called (see \cite{[M1]}) inverse
Gaussian.} $F(x;\muIG,\lambdaIG)$ is concentrated on the positive half-line;
its mean is $\muIG$, variance is ${\muIG^{\,3}}/{\lambdaIG}$, and the third
central moment is ${3\,\muIG^{\,5}}/{\lambdaIG^{2}}$. The appearance of this
skewed distribution in Theorem~\ref{esrytrf} sheds light on numerous claims by
many practitioners (see, e.g., \cite{[Floreani=2013]}, \cite{[Mittnik=2011]},
\cite{[Pfeifer=Strassburger=2008]}) that the ``world of normal or, more
generally, elliptically contoured risk distributions'' is chosen wrongly when
the solvency problems are considered, whereas the ``world of skewed
distributions'' is adequate in this framework.
\end{remark}

\begin{table}[t]
\caption{\small Models in Figs.~\ref{sdrtfrjmty}--\ref{srtyikyer}.}
\vglue -8pt
{\small\begin{tabular}{p{1.0cm}|p{3.7cm}|p{3.7cm}|p{1.1cm}|p{1.1cm}}
\hline\noalign{\smallskip}  & $T$ & $Y$ & $M$ & $D^{\,2}$
\\[0pt]
\noalign{\smallskip}\hline\hline\noalign{\smallskip} Fig.~\ref{sdrtfrjmty}: &
exponentially distributed; $\paramT=1$ & exponentially distributed; $\paramY=1$
& $1$ & $2$
\\[0pt]
\noalign{\smallskip}\hline\noalign{\smallskip}
Fig.~\ref{qwertyrjty}: & 2-mixture; & Pareto; & $0{.}8750$ & $2{.}3042$
\\[-4pt]
\noalign{\smallskip}\noalign{\smallskip}
& $\paramT_1=1$, $\paramT_2=2$, $p=2/3$ & $\parA{Y}=4{.}0$, $\parB{Y}=0{.}35$ &
&
\\[0pt]
\noalign{\smallskip}\hline\noalign{\smallskip}
Fig.~\ref{ftuytrjityy}: & Erlang;  & Pareto; & $0{.}8$ & $1{.}2$
\\[-4pt]
\noalign{\smallskip}\noalign{\smallskip}
& $\paramT=6{.}0$, $k=4$ & $\parA{Y}=4{.}0$, $\parB{Y}=0{.}4$ & &
\\[0pt]
\noalign{\smallskip}\hline\noalign{\smallskip}
Fig.~\ref{srtyikyer}: & Pareto; & Pareto; & $1$ & $1{.}3333$
\\[-4pt]
\noalign{\smallskip}\noalign{\smallskip}
&$\parA{T}=4{.}0$, $\parB{T}=0{.}4$ & $\parA{Y}=4{.}0$, $\parB{Y}=0{.}4$ & &
\\[0pt]
\noalign{\smallskip}\hline\noalign{\smallskip}
\end{tabular}}\label{rtyurtujrtt}
\end{table}

Conditions of Theorem~\ref{esrytrf} are very general for both $T$ and $Y$, and
the accuracy of approximation of $\P\{\RtimeR{u,c}\leqslant t\}$ by
$\AInt{M}{u,c}(t)$ is high for all $c$ for which these values are not
negligibly small; this is a satisfactory approximation for the left-hand side
of equation \eqref{wertyjhft} that defines $\Ruat(c)$, $c\geqslant 0$, as an
implicit function.

To demonstrate this in a spectacular way, we compare Figs.~\ref{ewrtyhrtjr} and
\ref{sdrtfrjmty}. This shows the difference in the accuracy of the normal (or
Cram{\'e}r's) and inverse Gaussian approximations. The advantages of the latter
are especially noticeable in that domain (including the point $\cS$ and its
neighborhood) where $\P\{\RtimeR{u,c}\leqslant t\}$ assumes not too small
values.

To emphasize that the inverse Gaussian approximation works well for
heavy-tailed $Y$, we address Figs.~\ref{qwertyrjty}--\ref{srtyikyer} (see
Table~\ref{rtyurtujrtt}), where $t=1000$, $u=40$. To get simulated values of
$\P\{\RtimeR{u,c}\leqslant t\}$ according to the algorithm described in
\cite{[M3]}, we take $\Span=0{.}05$ (and even less in the vicinity of $\cS$,
where this function's flexure is considerable), and $N=1000$.

In Fig.~\ref{qwertyrjty}, we draw the graph of $\AInt{M}{u,c}(t)$ calculated by
means of numerical integration in \eqref{srdfdsfgs} for $T$ 2-mixture of
exponential with parameters $\paramT_1=1$, $\paramT_2=2$, $p=2/3$ and $Y$
Pareto with parameters $\parA{Y}=4{.}0$, $\parB{Y}=0{.}35$, whence
$\cS=1{.}143$, $M=0{.}8750$, and $D^{\,2}=2{.}3042$. In Fig.~\ref{ftuytrjityy},
we do this for $T$ Erlang with parameters $\paramT=6{.}0$, $k=4$ and $Y$ Pareto
with parameters $\parA{Y}=4{.}0$, $\parB{Y}=0{.}4$, whence $\cS=1{.}25$,
$M=0{.}8$, and $D^{\,2}=1{.}2$. In Fig.~\ref{srtyikyer}, we do this for $T$
Pareto with parameters $\parA{T}=4{.}0$, $\parB{T}=0{.}4$ and $Y$ Pareto with
parameters $\parA{Y}=4{.}0$, $\parB{Y}=0{.}4$, whence $\cS=1$, $M=1$, and
$D^{\,2}=1.3333$.

\subsection{Non-ruin capital}\label{qwsrxthrjhrt}

The analytical technique which yields Theorem~\ref{esrytrf} (see
\cite{[Malinovskii=2018=dan=1]} and \cite{[M1]}, \cite{[M2]}, \cite{[M3]}) is
suitable for asymptotic analysis of non-ruin capital in the general risk model.
The following theorem (see \cite{[Malinovskii=2020=dan=2]}, Theorem~1) gives an
asymptotic representation for $\Ruat(c)$ at the points $c=0$ and $c=\cS$, where
(see \eqref{adsfgherfjnfnm}) $\cS=1/M$; this generalizes asymptotic equalities
\eqref{sfddfdbcsbb}.

\begin{theorem}\label{gyuiytik}
Assume that p.d.f. $f_{T}(x)$ and $f_{Y}(x)$ are bounded above by a finite
constant, $D^2>0$, $\E({T}^3)<\infty$, $\E({Y}^3)<\infty$. Then
\begin{equation*}
\begin{aligned}
\Ruat(0)&=\frac{t}{M}+\frac{D}{M^{\,3/2}}\,\NQuant{\alpha}\sqrt{t}\,(1+{o}(1)),\quad
t\to\infty,
\\[0pt]
\Ruat(\cS)&=\dfrac{D}{M^{\,3/2}}\,\NQuant{\alpha/2}\sqrt{t}\,(1+{o}(1)),\quad
t\to\infty.
\end{aligned}
\end{equation*}
\end{theorem}

The following theorem (see \cite{[Malinovskii=2020=dan=2]}, Theorem~2) gives an
asymptotic representation for $\Ruat(c)$, $c\geqslant 0$, which generalizes
asymptotic equality \eqref{dtyrkkjuy}.

\begin{theorem}\label{ergtrewghwerg}
Assume that p.d.f. $f_{T}(x)$ and $f_{Y}(x)$ are bounded above by a finite
constant, $D^2>0$, $\E({T}^3)<\infty$, $\E({Y}^3)<\infty$. Then
\begin{equation*}
\Ruat(c)=\begin{cases}
(\cS-c)\,t+\dfrac{D}{M^{3/2}}\,\funU{\alpha,t}\left(\dfrac{M^{\,3/2}(\cS-c)}{D}\sqrt{t}\,\right)\,\sqrt{t},
&0\leqslant c\leqslant\cS,
\\[8pt]
\dfrac{D}{M^{3/2}}\,\funU{\alpha,t}\left(\dfrac{M^{\,3/2}(\cS-c)}{D}\sqrt{t}\,\right)\,\sqrt{t},&c>\cS,
\end{cases}
\end{equation*}
where for $t$ sufficiently large the function $\funU{\alpha,t}(x)$,
$x\in\Rline$, is continuous, monotone increasing, as $x$ increases from
$-\infty$ to $0$, monotone decreasing, as $x$ increases from $0$ to $\infty$,
and such that
\begin{equation*}
\lim_{x\to-\infty}\funU{\alpha,t}(x)=0,\
\lim_{x\to\infty}\funU{\alpha,t}(x)=\NQuant{\alpha}
\end{equation*}
and $\funU{\alpha,t}(0)=\NQuant{\alpha/2}(1+{o}(1))$, $t\to\infty$.
\end{theorem}

Let us construct simple bounds for $\Ruat(c)$, $c\geqslant 0$. First,
Theorem~\ref{ergtrewghwerg} yields the following bilateral asymptotic bounds:
for $0\leqslant c\leqslant\cS$, we have
\begin{equation}\label{adsfgsdbsb}
\begin{aligned}
(\cS-c)\,t+\dfrac{D}{M^{\,3/2}}\,&\NQuant{\alpha}\sqrt{t}\,(1+{o}(1))\leqslant\Ruat(c)
\\[0pt]
&\leqslant(\cS-c)\,t+\dfrac{D}{M^{\,3/2}}\,\NQuant{\alpha/2}\sqrt{t}\,(1+{o}(1)),\quad
t\to\infty.
\end{aligned}
\end{equation}
Second, looking for upper bounds for $\Ruat(c)$, $c>\cS$, we note that for $c$
sufficiently larger than $\cS$, i.e., for $c>K\cS$ with $K>1$ sufficiently
large, $\Ruat(c)$ is finite regardless of $t$. Thus, we focus (see
Remark~\ref{adsfgherjhr}) on $\RuaI(c)$, $c>\cS$, which is a natural upper
bound for $\Ruat(c)$, or\footnote{In order to have more freedom of action,
especially for finding compact formulas.} on any sensible upper bound for
$\RuaI(c)$.

Bearing in mind the widely known theory (see, e.g.,
\cite{[Rolski=et=al.=1999]}) built for the ultimate ruin probability
$\P\{\RtimeR{u,c}<\infty\}$, let us focus on the following cases.

\subsubsection*{M{\rm(}i{\rm)}: exponential case}\label{we5uy6r6tk}

When $T$ and $Y$ are exponentially distributed with parameters $\paramT$ and
$\paramY$, we have $\cS=\paramT/\paramY$, $\adjustL=\paramY-\paramT/c$. For
$c>\paramT/\paramY$, we have (see, e.g., \cite{[Rolski=et=al.=1999]})
$\P\{\RtimeR{u,c}<\infty\}=(1-\adjustL/\paramY)\,e^{-\adjustL\,u}$ for all
$u\geqslant 0$. This rewrites as
\begin{equation*}
\P\{\RtimeR{u,c}<\infty\}=\left(\paramT/(c\paramY)\right)
\,\exp\left\{-(\paramY-\paramT/c)\,u\right\},\quad c>\paramT/\paramY,
\end{equation*}
and by simple calculations we have
\begin{equation}\label{sdrtydr}
\Ruat(c)\leqslant\max\left\{\,0,-\frac{\ln\left(\alpha
c\paramY/\paramT\right)}{\paramY-\paramT/c}\right\},\quad c>\paramT/\paramY.
\end{equation}

\subsubsection*{M{\rm(}ii{\rm)}: Poisson claims arrival and $Y$ light-tailed}\label{dsghjtmky}

When $T$ is exponentially distributed with parameter $\paramT$ and the
distribution of $Y$ is light-tailed, but non-exponential, special cases of
which are, e.g.,
\begin{itemize}
\item[(\emph{a})] $T$ exponentially distributed and $Y$ 2-mixture,
\item[(\emph{b})] $T$ exponentially distributed and $Y$ Erlang,
\end{itemize}
we have $\cS=\paramT\,\E{Y}$, equation \eqref{sdfgrhjrgde} rewrites as
$\E\exp\{\adjustL\,Y\}=1+c\adjustL/\paramT$, and $\adjustL$ is its positive
solution. For $c>\cS$, we have (see, e.g., \cite{[Rolski=et=al.=1999]})
$\P\{\RtimeR{u,c}<\infty\}\leqslant e^{-\adjustL\,u}$ for all $u\geqslant 0$.
Therefore, by simple calculations we have
\begin{equation*}
\Ruat(c)\leqslant-{\ln\alpha}/{\adjustL},\quad c>\paramT\,\E{Y},
\end{equation*}
and the problem comes down to finding $\adjustL$ in a closed form.

\subsubsection*{M{\rm(}iii{\rm)}: Poisson claims arrival and $Y$ heavy-tailed}\label{drtfytjk}

Special cases are, e.g.,
\begin{itemize}
\item[(\emph{a})] $T$ exponentially distributed and $Y$ Pareto,
\item[(\emph{b})] $T$ exponentially distributed and $Y$
Kummer\footnote{Definition of the Kummer distribution see, e.g., in
\cite{[Malinovskii=1998]}.}.
\end{itemize}
Any upper bounds for $\Ruat(c)$, $c>\cS$, which assumes \emph{small, rather
than large} values, is tightly related to particulars of the probability
$\P\{\RtimeR{u,c}<\infty\}$ for \emph{small, rather than large} values of $u$,
which is a problem beyond the scope of this article.

\subsubsection*{M{\rm(}iv{\rm)}: renewal claims arrival and $Y$ exponentially distributed}\label{edrtrthjr}

When $Y$ is exponentially distributed with parameter $\paramY$ and the
distribution of $T$ is arbitrary, special cases of which are, e.g.,
\begin{itemize}
\item[(\emph{a})] $T$ 2-mixture and $Y$ exponentially distributed,
\item[(\emph{b})] $T$ Erlang and $Y$ exponentially distributed,
\item[(\emph{c})] $T$ Pareto and $Y$ exponentially distributed,
\item[(\emph{d})] $T$ Kummer and $Y$ exponentially distributed,
\end{itemize}
we have $\cS=1/(\paramY\,\E{T})$, equation \eqref{sdfgrhjrgde} rewrites as
$\E\exp\{-\adjustL\,c\,T\}=1-\adjustL/\paramY$, and $\adjustL$ is its positive
solution. For $c>\cS$, we have
$\P\{\RtimeR{u,c}<\infty\}=(1-\adjustL/\paramY)\,e^{-\adjustL u}$ for all
$u\geqslant 0$. Bearing in mind that $1-\adjustL/\paramY\leqslant 1$, we have
(see \cite{[Rolski=et=al.=1999]}, Corollary~6.5.2)
\begin{equation}\label{we5434y321}
\Ruat(c)\leqslant-{\ln\alpha}/{\adjustL},\quad c>1/(\paramY\,\E{T}),
\end{equation}
and the problem comes down to finding $\adjustL$ in a closed form.

\subsubsection*{M{\rm(}v{\rm)}: renewal claims arrival and $Y$ light-tailed}\label{aserteru}

When $Y$ is light-tailed, but non-exponential, and the distribution of $T$ is
arbitrary, special cases of which are, e.g.,
\begin{itemize}
\item[(\emph{a})] $T$ Erlang and $Y$ Erlang,
\item[(\emph{b})] $T$ Erlang and $Y$ 2-mixture,
\end{itemize}
we address $X\overset{d}{=}Y-c\,T$, whose c.d.f. is $F_{X}$, denote by
$\overline{F}_{X}(x)=1-F_{X}(x)$ is tail function, and write
$x_{0}=\sup\{x:F_{X}(x)<1\}$. For $c>\cS={\E{Y}}/{\E{T}}$, we have (see
\cite{[Rolski=et=al.=1999]}, Theorem~6.5.4)
\begin{equation}\label{f67u56iu5}
b_{\circleddash}\,e^{-\adjustL
u}\leqslant\P\big\{\RtimeR{u,c}<\infty\big\}\leqslant b_{\oplus}\,e^{-\adjustL
u}\quad\text{for all}\quad u\geqslant 0,
\end{equation}
where $\adjustL$ is a positive solution to \eqref{sdfgrhjrgde} and
\begin{equation*}
b_{\oplus}=\inf_{x\in[0,x_{0}]}\frac{e^{\adjustL
x}\overline{F}_{X}(x)}{\int_{x}^{\infty}e^{\adjustL y}\,dF_{X}(y)},\quad
b_{\circleddash}=\sup_{x\in[0,x_{0}]}\frac{e^{\adjustL
x}\overline{F}_{X}(x)}{\int_{x}^{\infty}e^{\adjustL y}\,dF_{X}(y)}.
\end{equation*}
Alternatively (see \cite{[Rolski=et=al.=1999]}, Theorem~6.5.5), the
inequalities \eqref{f67u56iu5} hold with
\begin{equation*}
b_{\oplus}^{*}=\inf_{x\in[0,x_{0}^{*}]}\frac{e^{\adjustL
x}\overline{F}_{Y}(x)}{\int_{x}^{\infty}e^{\adjustL y}\,dF_{Y}(y)},\quad
b_{\circleddash}^{*}=\sup_{x\in[0,x_{0}^{*}]}\frac{e^{\adjustL
x}\overline{F}_{Y}(x)}{\int_{x}^{\infty}e^{\adjustL y}\,dF_{Y}(y)},
\end{equation*}
where $x_{0}^{*}=\sup\{x:F_{Y}(x)<1\}$; the inequalities $0\leqslant
b_{\circleddash}^{*}\leqslant b_{\circleddash}\leqslant b_{\oplus}\leqslant
b_{\oplus}^{*}\leqslant 1$ hold.

Both upper and lower bounds for $\RuaI(c)$, $c>\cS$, which is a solution to
equation \eqref{aesfgewge}, and therefore upper bounds for $\Ruat(c)$, $c>\cS$,
is easy to get from \eqref{f67u56iu5}, and we leave this to the reader.

\subsubsection*{M{\rm(}vi{\rm)}: renewal claims arrival and $Y$ heavy-tailed}\label{rftyhrtwe}

Special cases are, e.g.,
\begin{itemize}
\item[(\emph{a})] $T$ is 2-mixture and $Y$ is Pareto,
\item[(\emph{b})] $T$ is Erlang and $Y$ is Pareto,
\item[(\emph{c})] $T$ is Pareto and $Y$ is Pareto.
\end{itemize}
Any upper bounds for $\Ruat(c)$, $c>\cS$, which assumes \emph{small, rather
than large} values, is tightly related to particulars of the probability
$\P\{\RtimeR{u,c}<\infty\}$ for \emph{small, rather than large} values of $u$,
which is a problem beyond the scope of this article.

\section{Numerical illustrations of non-ruin capital's structure}\label{ewrtyh6jh5rt}

Let us compare numerically the results formulated in Section~\ref{qwsrxthrjhrt}
with the simulation results taken as exact values; the algorithm of simulation
is the same as in \cite{[Malinovskii=Kosova=2014]}, or in \cite{[M3]}. For
completeness, we return to Section~\ref{qewrtehr} and start with $T$ and $Y$
exponentially distributed with parameters $\paramT$ and $\paramY$, whose p.d.f.
are
\begin{equation*}
f_{T}(x)=\paramT\,e^{-\paramT x},\quad f_{Y}(x)=\paramY\,e^{-\paramY x},\quad
x>0.
\end{equation*}

\subsection{Model $\boldsymbol{M}${\bf(}$\boldsymbol{i}${\bf)}: exponential case}

Elementary calculations yield $\E(T^k)={k!}/{\paramT^{k}}$,
$\E(Y^k)={k!}/{\paramY^{k}}$, $k=1,2,\dots$, whence
\begin{equation*}
\begin{aligned}
\E{T}&={1}/{\paramT},& \D{T}&={1}/{\paramT^{\,2}},
\\[0pt]
\E{Y}&={1}/{\paramY}, & \D{Y}&=1/\paramY^{2},
\end{aligned}
\end{equation*}
and
\begin{equation*}
\begin{aligned}
\E{e^{-\adjustL\,c\,T}}&=\paramT\int_{0}^{\infty}e^{-(\adjustL
c+\paramT)x}\,dx={\paramT}/{(\paramT+c\adjustL)},
\\[0pt]
\E{e^{\adjustL\,Y}}&=\paramY\int_{0}^{\infty}e^{(\adjustL-\paramY)
x}\,dx={\paramY}/{(\paramY-\adjustL)}.
\end{aligned}
\end{equation*}
Plainly, $\cS=\E{Y}/\E{T}$ is equal to $\paramT/\paramY$, the constants defined
in \eqref{adsfgherfjnfnm} are
\begin{equation*}
\begin{aligned}
M&=\E{T}/\E{Y}={\paramY}/{\paramT},
\\[0pt]
D^{\,2}&=\big((\E{T})^{2}\D{Y}+(\E{Y})^{2}\D{T}\big)/(\E{Y})^{3}={2\,\paramY}/{\paramT^{\,2}},
\end{aligned}
\end{equation*}
and for $c>\paramT/\paramY$ the positive solution $\adjustL$ to the Lundberg
equation \eqref{sdfgrhjrgde}, which rewrites as the quadratic equation
$\big(\paramY-\adjustL\big)\,\big(\paramT+c\adjustL\big)-\paramT\paramY=0$, is
$\adjustL=\paramY-\paramT/c$.

\begin{figure}[t]
\center{\includegraphics[scale=.8]{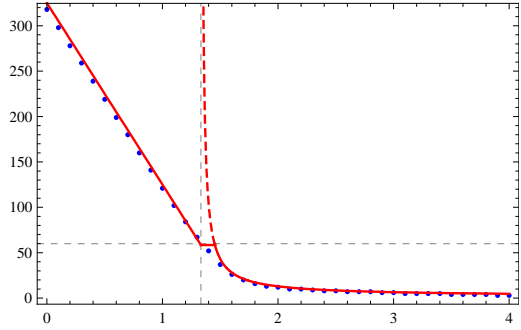}}
\caption{\small \emph{Model~$M(i)$}: upper bound ($X$-axis is $c$) on
$\Ruat(c)$ and simulated values of $\Ruat(c)$, drawn for $T$ and $Y$
exponentially distributed with parameters $\paramT=3/5$, $\paramY=4/5$, and
$\alpha=0{.}05$, $t=200$. Vertical grid line: $\cS=4/3$. Horizontal grid line:
$\Ruat(\cS)=59{.}9033$.}\label{stydfjhfgj}
\end{figure}

In Fig.~\ref{stydfjhfgj}, the upper bounds \eqref{adsfgsdbsb} in the case
$0\leqslant c\leqslant\paramT/\paramY$, and \eqref{sdrtydr} in the case
$c>\paramT/\paramY$, are drawn for $t=200$, $\alpha=0{.}05$, $\paramT=4/5$,
$\paramY=3/5$, whence $\cS=1{.}3333$, $M=0{.}75$, and $D^{\,2}=1{.}875$. In
Fig.~\ref{stydfjhfgj}, by dots are drawn the simulated values of $\Ruat(c)$.

\subsection{Model $\boldsymbol{M}${\bf(}$\boldsymbol{iv}${\bf)}: Erlang $\boldsymbol{T}$
and exponentially distributed $\boldsymbol{Y}$}

This model is a particular case of Model $M(v)$, where $T$ is Erlang with
parameters $k$ integer and $\paramT>0$ and $Y$ is Erlang with parameters $m$
integer and $\paramY>0$, whose p.d.f. are
\begin{equation*}
f_{T}(x)=\frac{\paramT^{\,k} x^{\,k-1}}{\Gamma(k)}\,e^{-\paramT x},\quad
f_{Y}(x)=\frac{\paramY^{\,m} x^{\,m-1}}{\Gamma(m)}\,e^{-\paramY x},\quad x>0.
\end{equation*}
Elementary calculations yield
\begin{equation*}
\begin{aligned}
\E{T}&=k/\paramT,& \D{T}&={k}/\paramT^{\,2},
\\[0pt]
\E{Y}&={m}/{\paramY},& \D{Y}&={m}/{\paramY^{\,2}},
\end{aligned}
\end{equation*}
and
\begin{equation*}
\begin{aligned}
\E{e^{-\adjustL\,c\,T}}&=\frac{\paramT^{\,k}}{\Gamma(k)}\int_{0}^{\infty}e^{-(\adjustL
c+\paramT)x}x^{\,k-1}\,dx=\frac{\paramT^{\,k}}{(\paramT+c\adjustL)^k},
\\[0pt]
\E{e^{\adjustL\,Y}}&=\frac{\paramY^{m}}{\Gamma(m)}\int_{0}^{\infty}e^{(\adjustL-\paramY)
x}x^{m-1}\,dx=\frac{\paramY^m}{(\paramY-\adjustL)^m}.
\end{aligned}
\end{equation*}
Plainly, $\cS=\E{Y}/\E{T}$ is equal to ${(m\paramT)}/{(k\paramY)}$, the
constants defined in \eqref{adsfgherfjnfnm} are
\begin{equation*}
\begin{aligned}
M&=\E{T}/\E{Y}={k\paramY}/{(m\paramT)},
\\[0pt]
D^{\,2}&=((\E{T})^{2}\D{Y}+(\E{Y})^{2}\D{T})/(\E{Y})^{3}
\\[0pt]
&=k\,(k+m)\,\paramY/(m^2\paramT^{\,2}),
\end{aligned}
\end{equation*}
and for $c>{(m\paramT)}/{(k\paramY)}$ the positive solution $\adjustL$ to the
Lundberg equation \eqref{sdfgrhjrgde}, which rewrites as
$(\paramY-\adjustL)^{m}\,(\paramT+c\adjustL)^{k}-\paramT^{\,k}\paramY^{\,m}=0$,
is easy to find numerically; this $\adjustL$ is not explicit, except for $m=1$.

\begin{figure}[t]
\center{\includegraphics[scale=.8]{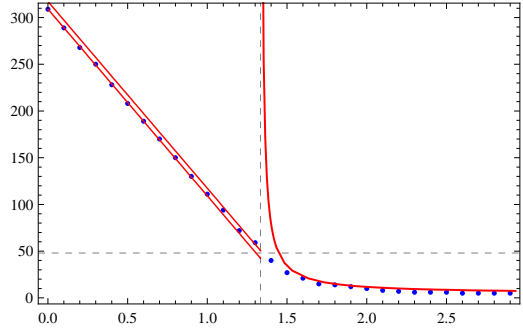}}
\caption{\small\emph{Model~$M(iv)$}: graphs ($X$-axis is $c$) of two-sided
bounds \eqref{adsfgsdbsb}, when $0\leqslant c\leqslant\cS$, of the upper bound,
when $c>\cS$, and of simulated values of $\Ruat(c)$, drawn for $T$ which is
Erlang with parameters $\paramT=8/5$, $k=2$ and $Y$ exponentially distributed
with parameter $\paramY=3/5$, and $\alpha=0{.}05$, $t=200$. Vertical grid line:
$\cS=4/3$. Horizontal grid line: $\Ruat(\cS)=48$.}\label{egrrgd}
\end{figure}

In Fig.~\ref{egrrgd}, the upper and lower bounds \eqref{adsfgsdbsb} in the case
$0\leqslant c\leqslant\paramT/\paramY$, and the upper bound \eqref{we5434y321}
in the case $c>\paramT/\paramY$, are drawn for $t=200$, $\alpha=0{.}05$,
$\paramT=8/5$, $k=2$, $\paramY=3/5$, whence $\cS=\E{Y}/\E{T}=1{.}3333$,
$M=0{.}75$, and $D^{\,2}=1{.}40625$. In Fig.~\ref{egrrgd}, by dots are drawn
the simulated values of $\Ruat(c)$.

\subsection{Model $\boldsymbol{M}${\bf(}$\boldsymbol{iii}${\bf)}: exponentially
distributed $\boldsymbol{T}$ and Pareto $\boldsymbol{Y}$}

\begin{figure}[t]
\center{\includegraphics[scale=.8]{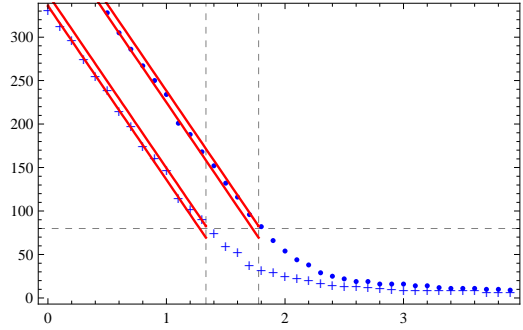}}
\caption{\small\emph{Model~$M(iii)$}: graphs ($X$-axis is $c$) of two-sided
bounds \eqref{adsfgsdbsb} for $0\leqslant c\leqslant\cS$ and of simulated
values of $\Ruat(c)$, drawn for $T$ exponentially distributed with parameter
$\paramT=4/5$ and $Y$ which is Pareto with parameters $\parA{Y}=10$,
$\parB{Y}=0{.}05$ (dots), $\parA{Y}=3$, $\parB{Y}=0{.}3$ (crosses), and
$\alpha=0{.}05$, $t=200$. Vertical grid lines: $\cS=1{.}7778$ (dots) and
$\cS=1{.}3333$ (crosses). Horizontal grid line: $\Ruat(\cS)=80$ (the same for
dots and crosses).}\label{hujhkugf}
\end{figure}

For $T$ exponentially distributed with parameter $\paramT>0$ and $Y$ whose
distribution is Pareto with parameters $\parA{Y}>0$, $\parB{Y}>0$, p.d.f. are
\begin{equation*}
f_{T}(x)=\paramT\,e^{-\paramT\,x},\quad
f_{Y}(x)=\dfrac{\parA{Y}\parB{Y}}{(x\,\parB{Y}+1)^{\parA{Y}+1}},\quad x>0,
\end{equation*}
elementary calculations yield
\begin{equation*}
\begin{aligned}
\E{T}&={1}/{\paramT},& \D{T}&={1}/{\paramT^{\,2}},
\\[0pt]
\E{Y}&={1}/{((\parA{Y}-1)\,\parB{Y})}, &
\D{Y}&={\parA{Y}}/{((\parA{Y}-1)^{2}\,(\parA{Y}-2)\,\parB{Y}^{2})}.
\end{aligned}
\end{equation*}
Plainly, $\cS=\E{Y}/\E{T}$ is equal to ${\paramT}/{((\parA{Y}-1)\,\parB{Y})}$,
the constants defined in \eqref{adsfgherfjnfnm} are
\begin{equation*}
\begin{aligned}
M&=\E{T}/\E{Y}=\frac{(\parA{Y}-1)\,\parB{Y}}{\paramT},
\\[0pt]
D^{\,2}&=((\E{T})^{2}\D{Y}+(\E{Y})^{2}\D{T})/(\E{Y})^{3}
=\frac{2\,(\parA{Y}-1)^{2}\,\parB{Y}}{\paramT^2\,(\parA{Y}-2)},
\end{aligned}
\end{equation*}
and the adjustment coefficient does not exist.

In Fig.~\ref{hujhkugf}, the upper and lower bounds \eqref{adsfgsdbsb} in the
case $0\leqslant c\leqslant\cS$ are drawn. Bounds for $c>\cS$ are beyond the
scope of this article and are not considered, although the essence of the
complexity in their construction is clear. By dots, drawn are simulated values
of $\Ruat(c)$, $c\geqslant 0$. We note that for  $\parA{Y}=3$, $\parB{Y}=0{.}3$
(crosses), the third moment $\E({Y}^3)$ is not finite, and Fig.~\ref{hujhkugf}
suggests that the moment conditions in Theorems~\ref{gyuiytik} and
\ref{ergtrewghwerg} may be somewhat relaxed. However, this will significantly
complicate the proof, which lies beyond the scope of this article.

\subsection{Model $\boldsymbol{M}${\bf(}$\boldsymbol{iii}${\bf)}: exponentially
distributed $\boldsymbol{T}$ and Kummer $\boldsymbol{Y}$}

\begin{figure}[t]
\center{\includegraphics[scale=.8]{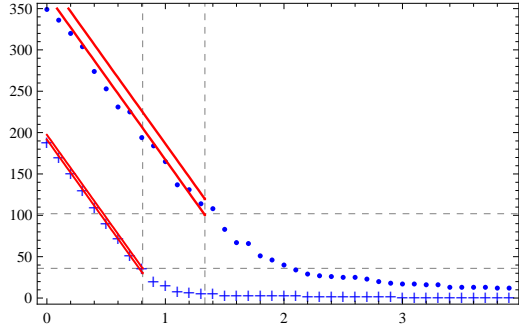}}
\caption{\small\emph{Model~$M(iii)$}: graph ($X$-axis is $c$) of two-sided
bounds \eqref{adsfgsdbsb} for $0\leqslant c\leqslant\cS$ and of simulated
values of $\Ruat(c)$, drawn for $T$ exponentially distributed with parameter
$\paramT=4/5$ and $Y$ which is Kummer with parameters $\kumK{Y}=5$,
$\kumL{Y}=5$ (dots), $\kumK{Y}=200$, $\kumL{Y}=200$ (crosses), $\alpha=0{.}05$,
and $t=200$. Vertical grid line: $\cS=1{.}3333$ (dots) and $\cS=0{.}8081$
(crosses). Horizontal grid lines: simulated $\Ruat(\cS)=102$ (dots) and
$\Ruat(\cS)=36$ (crosses).}\label{sfdgdbgfdbg}
\end{figure}

For $T$ exponentially distributed with parameter $\paramT>0$ and $Y$ whose
distribution is Kummer with parameters $\kumK{Y}>0$, $\kumL{Y}>0$, p.d.f.
are\footnote{For other equivalent formulas for $f_{Y}(x)$ see
\cite{[Malinovskii=1998]}.}
\begin{equation*}
f_{T}(x)=\paramT\,e^{-\paramT\,x},\quad
f_{Y}(x)=\frac{\kumK{Y}}{2}\frac{\Gamma\left(\frac{\kumK{Y}+\kumL{Y}}{2}\right)}
{\Gamma\left(\frac{\kumK{Y}}{2}\right)}\,\Kummer
\left(1+\frac{\kumL{Y}}{2},2-\frac{\kumK{Y}}{2},\frac{\kumK{Y}}{\kumL{Y}}\,x\,\right),\quad
x>0.
\end{equation*}
Elementary calculations yield
\begin{equation*}
\E{T^k}={k!}/{\paramT^{k}},\quad \E{Y^k}=
\dfrac{\Gamma\left(\frac{\kumK{Y}}{2}+k\right)\Gamma\left(\frac{\kumL{Y}}{2}-k\right)}
{\Gamma\left(\frac{\kumK{Y}}{2}\right)\Gamma\left(\frac{\kumL{Y}}{2}\right)}\,
\,\kumL{Y}^{\,k}\,\kumK{Y}^{-k},\quad
2k<\kumL{Y},\quad k=1,2,\dots.
\end{equation*}
In particular,
\begin{equation*}
\begin{aligned}
\E{T}&={1}/{\paramT},& \D{T}&={1}/{\paramT^{\,2}},
\\[0pt]
\E{Y}&=\dfrac{\kumL{Y}}{\kumL{Y}-2}, &
\D{Y}&=\dfrac{\kumL{Y}^{\,2}(\,4(\,\kumL{Y}-2)+\kumK{Y}\kumL{Y})}
{\kumK{Y}(\,\kumL{Y}-2)^{\,2}(\,\kumL{Y}-4)}.
\end{aligned}
\end{equation*}
Plainly, $\cS=\E{Y}/\E{T}$ is equal to ${\paramT\,\kumL{Y}}/{(\,\kumL{Y}-2)}$,
the constants defined in \eqref{adsfgherfjnfnm} are
\begin{equation*}
\begin{aligned}
M&=\E{T}/\E{Y}=\frac{(\,\kumL{Y}-2)}{\paramT\,\kumL{Y}},
\\[-2pt]
D^{\,2}&=\big((\E{T})^{2}\D{Y}+(\E{Y})^{2}\D{T}\big)/(\E{Y})^{3}
=\frac{2\,(2+\kumK{Y})(\,\kumL{Y}-2)^2}{\paramT^2
\kumK{Y}(\,\kumL{Y}-4)\,\kumL{Y}},
\end{aligned}
\end{equation*}
and the adjustment coefficient does not exist.

In Fig.~\ref{sfdgdbgfdbg}, the upper and lower bounds \eqref{adsfgsdbsb} in the
case $0\leqslant c\leqslant\cS$ are drawn. Bounds for $c>\cS$ are beyond the
scope of this article and are not considered, although the essence of the
complexity in their construction is clear. By dots, drawn are the simulated
values of $\Ruat(c)$, $c\geqslant 0$.

\section{Conclusion}\label{sartg54yh}

This paper provides a mathematical investigation of an observation (see, e.g.,
\cite{[Floreani=2013]}) made empirically, that the Value-at-Risk is not a good
solution to the problem of risk measure's choice balanced for efficiency and
simplicity. Regarding efficiency, the Value-at-Risk is not a good substitute
for non-ruin capital, which can be seen even from definitions. Regarding
simplicity, it turns out in fairly general risk models that the structure of
non-ruin capital is as simple as the structure of Value-at-Risk.

\end{document}